\documentclass[12pt]{article}
\usepackage{amsmath, amssymb, amsthm}
\usepackage{graphicx}
\usepackage{enumerate}
\usepackage{natbib}
\usepackage{url} 
\usepackage[utf8]{inputenc}
\usepackage{multirow}
\usepackage{algorithm}
\usepackage{algorithmic}
\usepackage{float}
\usepackage{caption}
\usepackage{subcaption}
\usepackage{booktabs}
\usepackage{mathtools}
\usepackage{xcolor}
\usepackage{titlesec}
\usepackage{hyperref}

\DeclareMathOperator{\E}{\mathbb{E}}

\newcommand{\blind}{1}

\begin{document}

\def\spacingset#1{\renewcommand{\baselinestretch}%
{#1}\small\normalsize} \spacingset{1}


\if1\blind
{
  \title{\bf Development of Public Health Policy by Digital Twin Microsimulation and Q-learning: A COVID-19 Booster Case Study\thanks{The paper has been accepted by the \textit{Journal of Americal Statistical Association}}}
  \author{Guoxuan Ma$^1$, Sicong Xie$^2$, Lili Zhao$^{3\dag}$ and Jian Kang$^{1}$\thanks{To whom correspondence should be addressed: zhaolili@northwestern.edu and jiankang@umich.edu}\\ Department of Biostatistics, University of Michigan$^1$\\Department of Statistics, University of Michigan$^2$\\ Feinberg School of Medicine, Northwestern University$^3$}  
  \maketitle
} \fi

\if0\blind
{
  \bigskip
  \bigskip
  \bigskip
  \begin{center}
    \LARGE\bf Development of Public Health Policy by Digital Twin Microsimulation and Q-learning: A COVID-19 Booster Case Study
\end{center}
  \medskip
} \fi

\bigskip
\begin{abstract}
The COVID-19 pandemic highlighted the urgent need for effective vaccine policies, but traditional clinical trials often lack sufficient data to capture the diverse population characteristics necessary for comprehensive public health strategies. Ethical concerns around randomized trials during a pandemic further complicate policy development for public health. Reinforcement Learning (RL) offers a promising alternative for vaccine policy development. However, direct online RL exploration in real-world scenarios can result in suboptimal and potentially harmful decisions. This study proposes a novel framework combining tabular Q-learning with microsimulation, where a Recurrent Neural Network (RNN) serves as a digital twin environment simulator of the target population. This digital twin captures temporal associations between infection and patient characteristics to generate realistic individual disease trajectories, enabling safe and efficient policy learning without real-world interaction. Our tabular Q-learning model produces an interpretable policy table that balances the risks of severe infection against vaccination side effects. Applied to COVID-19 booster policies, the learned Q-learning-based policy outperforms current practices, offering a path toward more effective vaccination strategies. A project webpage introducing our work, including links to the software, a brief introductory video, and a step-by-step tutorial video, is available at \href{https://public.websites.umich.edu/~jiankang/software/dtpl_website_umich/index.html}{\url{https://public.websites.umich.edu/\~jiankang/software/dtpl\_website\_umich/index.html}}.
\end{abstract}

\noindent%
\vfill

\newpage
\spacingset{1.8}

\section{Introduction}

The COVID-19 pandemic underscored the critical importance of rapid and effective vaccination strategies to control the spread of the virus and minimize the burden on healthcare systems. However, developing optimal vaccination policies for public health is fraught with challenges. Clinical trials do not have sufficient data to evaluate vaccine policies comprehensively, as they often enroll subjects with specific characteristics that may not be representative of the general population \citep{juni2001assessing, lander2019involving}. As a result, the findings from these trials may not always generalize well to the broader population, potentially hindering the development of comprehensive vaccination policies. For instance, people taking immunosuppressant medications were excluded from the trials developing BNT162b2 (Pfizer-BioNTech) and mRNA-1273 (Moderna) vaccines \citep{polack2020safety, baden2021efficacy}. The lack of data in this group regarding the vaccines efficacy  has prevented the development of an effective vaccine policy against COVID-19 infections for immunosuppressed patients \citep{risk2022covid}. Moreover, conducting large-scale randomized trials on vaccine policy evaluation during a pandemic poses significant ethical challenges. Randomizing subjects to the group that do not receive the vaccine can place participants at increased risk of COVID-19 infection, raising ethical concerns about exposing groups of people to potential harm \citep{adebamowo2014randomised, monrad2020ethical}. These challenges highlight the need for alternative approaches to develop and improve vaccine policies using existing real-world data while adhering to safety and ethical standards.

The electronic health record (EHR) dataset from University of Michigan Hospital and Michigan Medicine (study ID: HUM00164771), which spans both the pre-pandemic and pandemic periods, provides a comprehensive resource for studying COVID-19 vaccination strategies in a real-world context. This dataset comprises extensive health records from 1,224,147 patients, including demographics (e.g., age, gender, and race), COVID-19 vaccination dates, infection history, and other baseline characteristics such as immunosuppressant usage, number of hospital visits, and comorbidities. The breadth and granularity of this dataset enable a systematic investigation of vaccination policies that account for heterogeneity across patient populations. Based on this EHR dataset, we aim to develop vaccination policies that improve upon current observed practices and better address the needs of diverse patient populations, including the immunosuppressant groups that were underrepresented in clinical trials \citep{polack2020safety, risk2022covid}. 

Reinforcement learning (RL) offers a promising framework for developing adaptive vaccination strategies and has been successfully applied in various fields in healthcare \citep{sutton2018reinforcement, yu2021reinforcement}, including cancer treatment, glucose regulation, HIV treatment, and mental diseases intervention \citep{tseng2017deep, sun2018dual, yu2019incorporating, laber2014interactive}, but it has not been widely adopted in public health \citep{weltz2022reinforcement}. In an RL setup, an agent selects actions based on its current state, receiving feedback (rewards) and the new state from the environment. The objective is for the agent to learn an optimal policy, which is a mapping from states to actions, that maximizes the cumulative reward over time. RL is particularly suited for systems with inherent delays, where decisions are made sequentially without immediate feedback and are evaluated based on long-term outcomes. This makes RL a compelling approach for developing effective policies in public health, including policymaking in vaccination, where the health outcomes are often evaluated based on a prolonged period with delayed feedback \citep{yu2021reinforcement}.

However, RL agents do not receive explicit instructions on which actions to take; instead, they learn the best actions through trial and error (in the online setting) or learn from the existing data (in the offline setting). While the online trial-and-error process encourages agents to explore new policies that are potentially effective, applying this approach directly in real-world scenarios can raise ethical concerns \citep{levine2020offline}. Early in the training process, the trial-and-error learning mechanism often lead to suboptimal or even harmful decisions, potentially causing harm to subjects before corrective feedback is obtained. While this may be less problematic in applying RL in Dynamic Treatment Regimes \citep{liu2017deep, zhang2020designing, guo2022learning}, where treatment decisions are usually made under experts' supervision to ensure they are clinically relevant and safe, it becomes more critical in scenarios of learning vaccine policy during a global pandemic like COVID-19, where it is impossible to provide individualized supervision for the whole population. In such cases, incorrect vaccination timing or administration could lead to severe infections or serious adverse effects. This problem in online training can be resolved by an offline approach, where actions are learned based on observed data. However, the offline approach may struggle with exploring new policies that could lead to potential improvements and cannot effectively learn from the observed offline data \citep{levine2020offline}. Moreover, it faces the same ethical concerns when evaluating the learned policy in real world with the absence of proper supervisions. Therefore, there is a need to develop an online RL framework based on existing data without direct interactions with the real world.

Q-learning is a Reinforcement Learning algorithm that helps agents learn how to make decisions by evaluating the potential value of different actions in various states \citep{watkins1992q}. It maintains a Q-function of the state-action pairs, which represents the expected future reward of taking a particular action in a given state. To determine the best action for a given state, the agent examines the Q-values associated with all possible actions and selects the one with the highest value. The Q-value for each state-action pair is initially designed to be stored explicitly in a table, as seen in tabular Q-learning \citep{watkins1992q}. Later, deep neural networks (DNN) are often used to model the Q-function for its flexibility and ability to accommodate continuous state and action spaces \citep{mnih2015human, liu2017deep, yu2021reinforcement}. While deep Q-learning is useful in precision medicine or individualized treatment where state and action spaces are often large or continuous \citep{liu2017deep, zhang2020designing, guo2022learning, wu2023value}, public health problems can be effectively represented with a finite and discrete set of states and actions, policies are often applied to groups of people rather than being tailored to individual subjects. Moreover, unlike tabular Q-learning, deep Q-learning lacks theoretical convergence guarantees to the global optimum due to the complexities of neural networks and function approximation and often suffers from convergence difficulties \citep{watkins1992q, van2016deep, fan2020theoretical}.

Many efforts have focused on estimating the impact of government interventions on the epidemiological spread of COVID-19. For instance, \citet{chernozhukov2020causal} utilized a causal structural model to investigate the effects of policies on COVID-19 growth and mortality rates. \citet{eftekhari2020markovian} compared Markovian and non-Markovian processes for lockdown allocation, while \citet{tian2021effects} employed synthetic control, discontinuity regression and state-space compartmental models to evaluate intervention stringency. \citet{tan2022transmission} employed compartmental models to estimate transmissibility of symptomatic and asymptomatic cases. Other approaches relied on RL for government-level policy optimization for COVID-19. \citet{kompella2020reinforcement} proposed an RL-based framework for fine-grained mitigation, and \citet{wan2021multi} applied RL to minimize long-term societal costs, both using epidemiological models as the environment simulator. However, these methods primarily address population-level infection dynamics or spatial spread. Because these simulators do not generate individual-level trajectories, they are unsuitable for developing vaccination policies tailored to specific subgroups. While \citet{kerr2021covasim} took an agent-based approach to projecting epidemic trends by simulating the state of individual people over discrete time points, their framework does not address sequential decision-making optimization required for targeted vaccination strategies.

\begin{figure}[t]
    \centering
    \small
    \spacingset{1}
    \includegraphics[scale = 0.1]{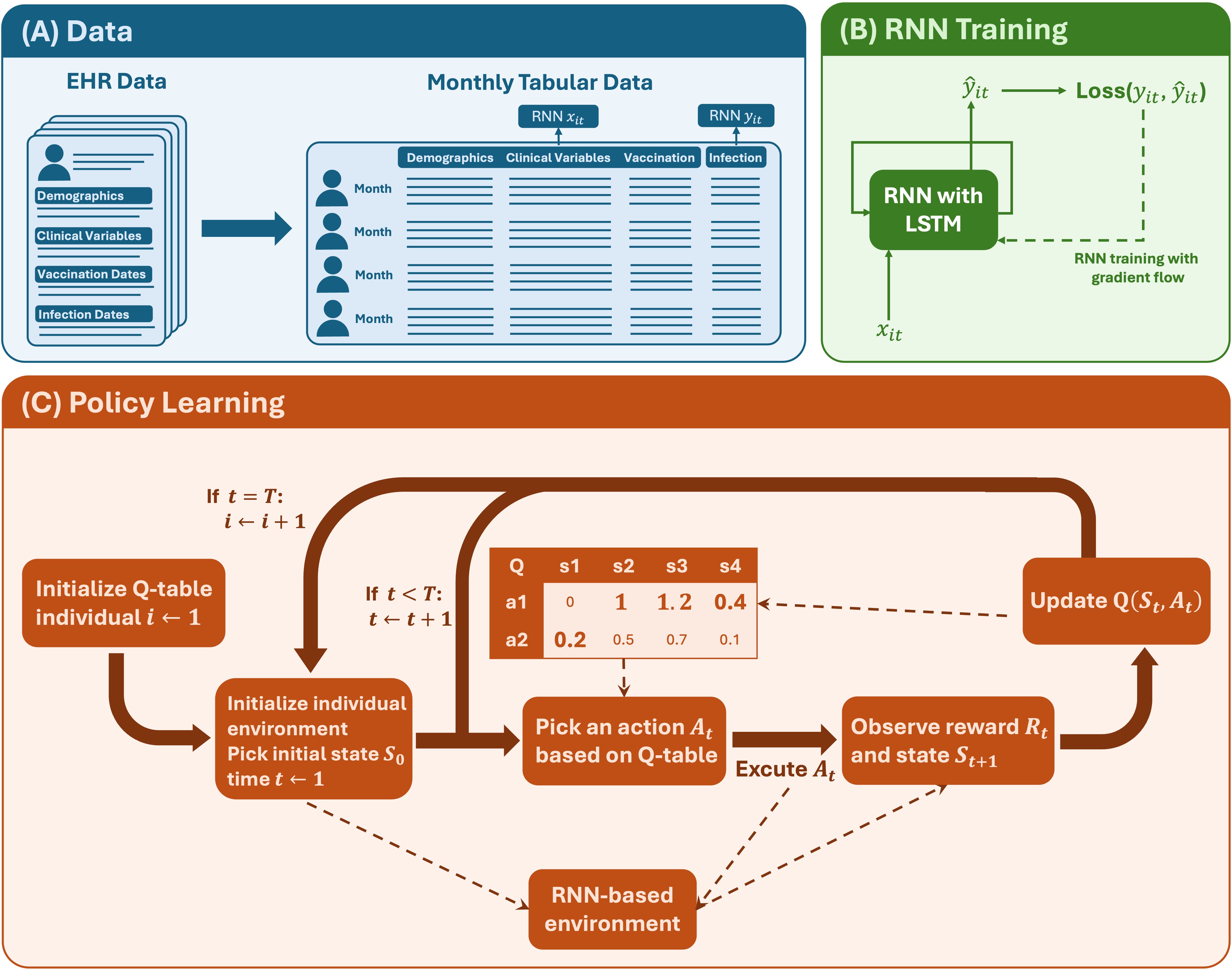}
    \caption{Our policy learning framework consisted of digital twin microsimulation and Q-learning. (A) An illustration of the data. The EHR data, including demographics, clinical variables, vaccination and infection dates, is converted to monthly tabular data for RNN training. For individual $i$, $x_{it}$ represents RNN predictors and $y_{it}$ represents RNN outcomes, for $t=1,\ldots,T$. Section \ref{sec:method_microsimulation} includes details on variables included in $x_{it}$ and $y_{it}$. (B) Training an RNN with LSTM architecture using monthly tabular data. (C) Steps of online tabular Q-learning where the environment is based on the fully-trained RNN with LSTM architecture from (B). Algorithm S1 provides more details on Q-learning steps.}
    \label{fig:model}
\end{figure}

To address these challenges, we propose a framework combining online tabular Q-learning with an RNN-based environment that interacts with the Q-learning agent. We refer to this RNN-based environment as a digital twin microsimulation model \citep{rutter2011dynamic, julien2022effect}, as it effectively simulates individual trajectories and creates a virtual environment that mirrors real-world dynamics. Figure \ref{fig:model} provides an overview. To illustrate the proposed framework, we focus on the development of COVID-19 vaccine policies in Michigan, using data collected from the University of Michigan Hospital. Nonetheless, the framework itself is broadly applicable to a wide range of sequential decision-making problems in public health and can be adapted to other populations and applications. We train an RNN with Long Short-Term Memory (LSTM) architecture, which has been widely used to model sequential data and has successful applications on modeling the relationship between COVID-19 infections and vaccinations \citep{sherstinsky2020fundamentals, shen2024state}, to capture the complex temporal association between infection status and patients' characteristics. Then, we perform online Q-learning, where the Q-function is modeled as a table, for vaccine policy learning based on the RNN microsimulator to avoid the need of actual executing the decision on whether to receive the vaccination. 

Our approach has two key contributions. First, by employing an RNN with the LSTM architecture, we create a digital twin microsimulation model as the virtual environment used in RL that can generate data for individuals that closely resembles real-world data. 
The RNN-based microsimulator not only avoids ethical concerns but also provides an unlimited amount of data for training, which creates ``what if'' scenarios, allowing us to evaluate different policies on simulated individuals with the same characteristics during the same period of time. Second, by using the tabular Q-learning, our approach produces a clear and interpretable policy table where Q-values for different actions and various groups of people can be easily read. We define the reward function by balancing the risk of severe infections and the potential side effects of the vaccination. In this paper, we focus on the vaccination policy for the COVID-19 booster dose, i.e., whether and when different groups of people should receive the booster after the second COVID-19 vaccination. The policy derived from our Q-table demonstrates superior performance compared to the current observed policy, indicating significant potential for improvement if the learned policy is adopted. A project webpage introducing our work, including links to the software, a brief introductory video, and a step-by-step tutorial video, is available at \href{https://public.websites.umich.edu/~jiankang/software/dtpl_website_umich/index.html}{\url{https://public.websites.umich.edu/~jiankang/software/dtpl\_website\_umich/index.html}}.

\section{Methods}

\subsection{Data processing}
We use deidentified electronic health record (EHR) dataset from 
the University of Michigan Hospital and Michigan Medicine (study ID: HUM00164771). The use of the data was approved by the Institutional Review Board (IRB) at the University of Michigan.
We include patients with a primary care physician and received at least one COVID-19 test at the University of Michigan Hospital. The dataset includes demographic variables, such as age, gender, and race, as well as baseline health measures, including the number of prior hospital visits and Charlson comorbidity index \citep{charlson1987new, gasparini2018comorbidity}. In addition, it records time-varying variables on patients’ COVID-19 infection and vaccination status. We exclude patients with race and gender missing as both variables are included in the microsimulation model. Since fewer than 1.5\% patients had either variables missing, we believe the exclusion is unlikely to introduce bias. A total of 81,000 patients are included in this study. The primary outcome we consider is whether the patient has severe COVID-19 infection (requiring hospitalization), and the secondary outcome is whether the patient has general COVID-19 infection. In this study, we summarize monthly data (illustrated in Figure \ref{fig:model}A) from the original EHR data starting from March 2020, and a patient's record terminates either at June 2022, the month of severe infection or decease, whichever comes first. The maximum span of data sequence for a patient is $T=27$ months, from March 2020 to June 2022, with minimum of 1 month, mean of 26.6 months, and median of 27 months. Although WHO did not declare the end of COVID-19 as a global health emergency until May 2023, 
the pandemic had largely subsided in the United States by mid-2022 \citep{CDC2024}. 
In addition, our dataset was available only through the end of June 2022 when we started the research. For these reasons, we selected June 2022 as the termination date. 

\subsection{Overview}

Let $\{S_t\}_{t\ge0}$ be a Markov process with state space $\mathcal{S}$ representing an individual's baseline characteristics, vaccination history and severe infection status at each month. Let $A_t$ be a random variable that represents the choice to receive a booster or not in month $t$ with action space $\mathcal{A}=\{0, 1\}$. For $s, s'\in\mathcal{S}$ and $a\in\mathcal{A}$, denote by $P(s'|s, a)= \Pr(S_{t+1} = s'\mid S_t = s, A_t = a)$ the state transition probability function. Let $\mathcal{R}(s, a)$ be a stationary reward function of taking action $a$ in state $s$. Let $R_t=\mathcal{R}(S_t, A_t)$, which is a random variable representing the reward at time $t$ after taking action $A_t$ at state $S_t$. Let $s_t$, $a_t$ and $r_t$ be the realizations of $S_t$, $A_t$ and $R_t$ respectively. Then, we model the decision-making process of whether an individual should receive a booster at each month as a Markov Decision Process (MDP), denoted by $\mathcal{M} = \{\mathcal{S}, \mathcal{A}, P, \mathcal{R}\}$. In this framework, the current state adequately summarizes the past, meaning future states are independent of past states given the current state and action. This assumption entails little loss of generality mathematically because almost any decision process can be reformulated as an MDP by appropriately aggregating the historical data \citep{sutton1997significance, weltz2022reinforcement}.

In Reinforcement Learning (RL), an agent interacts with the environment to improve its decision-making ability over time. An agent is a decision-maker responsible for determine whether an individual should receive a COVID-19 booster shot. The agent selects an action based on the current state according to a policy $\pi:\mathcal{S}\mapsto\mathcal{A}$, which  maps a patient's health status, vaccination history, and other relevant information (a state) to an action (whether to administer a booster or not). The environment, on the other hand,  consists of the state transition function $P(s'|s, a)$ and the reward function $\mathcal{R}(s, a)$. The environment interacts with the agent by providing the next state and reward after each action is taken.

We first introduce the Q-learning framework to derive a good policy for booster vaccination, assuming the environment is known. However, due to ethical concerns around allowing an agent to directly interact with real-world patients in the development of booster polices, we construct a microsimulation model as a virtual environment using RNN trained on existing patient data. Algorithm S1 outlines the steps of online tabular Q-learning with the RNN-based environment in our application of booster policy development. 

\subsection{Booster policy learning by tabular Q-learning}

We perform online tabular Q-learning given a known environment. The objective is to find a policy $\pi^*$ maximizing the expected cumulative reward, defined by the value function,
 $   V^\pi(s_0) = \E\left[\sum_{t=0}^{\infty}\gamma^t R_t \mid S_0 = s_0\right],\nonumber$
where $V^\pi(s_0)$ represents the expected cumulative reward starting from an initial state $s_0$ by following a policy $\pi$, and $\gamma\in(0, 1]$ is the discount factor on future reward. The expectation is taken over possible trajectories of the Markov process $\{S_t\}_{t\ge0}$ generated by the policy $\pi$ starting from the initial state $s_0$. Because the MDP $\mathcal{M}$ is defined with a stationary transition kernel $P(s'|s, a)$, the value function is also stationary. This means for any state, its value remains constant over time, as the recursive nature of the decision-making process ensures the value depends only on the current state, not on when it is encountered. Under the MDP framework, the optimal policy $\pi^*$ satisfies the Bellman equation,
    $V^*(s) = \max_{\pi} \E\left\{\mathcal{R}\left(s, \pi(s)\right) + \gamma V^*(s') \right\}$, 
where $s$ is the current state and $s'$ is the next state. Instead of working directly with $V^*(s)$, Q-learning focuses on the optimal action–value function $q^*$, which satisfies its own Bellman optimality equation,
 $   q^*(s, a) = \mathbb{E}\left\{\mathcal{R}(s, a) + \gamma \max_{a'}q^*(s', a')\right\}, \nonumber$
where $q^*(s, a)$ represents the expected cumulative reward of taking action $a$ in state $s$ following the optimal policy thereafter. The optimal policy is then given by $\pi^*(s)=\arg\max_{a} q^*(s, a)$. Since the Q-function is initially unknown, we model it as a table, where each cell $(s, a)$ holds the estimated value of $q(s, a)$ \citep{watkins1992q}. The Q-learning algorithm takes an iterative approach to updating the Q-table: after each interaction with the environment, we adjust $q(s, a)$ based on the observed reward $r$ and the estimated future value by the following updating rule \citep{watkins1992q},
\begin{align}
    q(s, a)\leftarrow q(s, a) + \beta\left\{r + \gamma\max_u q(s', u) - q(s, a)\right\} \label{eq:tabular_q_update}
\end{align}
where $\beta$ is a prespecified learning rate and $s'$ is the next state.

In this study, we aim to learn a policy on whether and when to receive a COVID-19 booster, so we only consider subjects with at least two COVID-19 vaccinations and their trajectories after their second vaccinations. At any month $t$, we decide the state $S_t\in\mathcal{S}$ consists of four relevant variables: age (categorical, 18-29/30-49/50-64/65+), baseline immunosuppressant usage (binary), months since the last vaccination (categorical, 0-4/5-6/7+), and the severe infection status (binary). The action $A_t\in\mathcal{A}=\{0, 1\}$ indicates whether or not a booster is received. In this study, we follow the Centers for Disease Control and Prevention (CDC) guidelines that an additional COVID-19 vaccine should be at least 4 months following the previous dose. The guideline was for adult ages 65 years and older but we generalize it to all age groups for simplicity. Following this guideline, $A_t$ is constrained to be 0 regardless of the values in Q-table for $t$ within 4 months of the second vaccination. 

\subsection{Creating the environment through microsimulation}\label{sec:method_microsimulation}

The RL environment is characterized by two key components: the state transition rules and the reward function. We simulate state transitions using RNN, enabling the microsimulation of individual trajectories. The reward function is designed to balance the risk of severe infection with potential adverse effects of a COVID-19 booster.

\paragraph{Microsimulation of state transitions} Let $W_t$ represent a set of additional variables relevant to the individual's profile, though of less interest to the vaccine policy decision and not included in the state variable $S_t$. It includes an individual's baseline and time-varying characteristics. In this study, baseline characteristics include gender (Female/Male), race (categorical, Caucasian/African American/Others), baseline number of hospital visits (categorical, 0-4/5-9/10-19/20-49/50+), and baseline Charlson comorbidity (categorical, 0/1-2/3-4/5+). The baseline period is one year before the study period starts. 
Time-varying variables include COVID-19 variant (categorical, Alpha/Delta/Omicron) and total number of vaccinations received up to month $t$ (integer, 0-4). 

We train an RNN with LSTM units to approximate the transition dynamics between states. Conceptually, for month $t=1,\ldots,T-1$, let $X_t=[S_t, W_t, A_t]$ denote the RNN predictors and $Y_{t+1}=[S_{t+1}, W_{t+1}]$ denote the RNN outcomes. The RNN takes as input the sequence of predictors $X_1,\ldots,X_t$  and predicts the outcome $Y_{t+1}$. The transition dynamics from $[S_t, W_t]$ to $[S_{t+1}, W_{t+1}]$ given action $A_t$ are modeled by the fully trained RNN, from where we obtain the transition dynamics from $S_t$ to $S_{t+1}$. 

In this study, certain state variables, including age, baseline immunosuppressant usage and months since last vaccination, as well as the set of additional variables $W_t$ are deterministic given action and time. Therefore, these variables are excluded from the outcomes, as their transitions are fixed. Additionally, since severe infections are treated as a terminal event (i.e., the trajectory is terminated upon a severe infection), it is always zero in $S_t$ for month $t$ before termination. As the general infection is an important variable that affects the likelihood of severe infection, we include the binary general infection status besides the binary severe infection status into the outcome variable. The output layer of the RNN uses a sigmoid activation to predict the probability of infections in the next month. This allows us to sample transitions from $S_t$ to $S_{t+1}$ based on the underlying state transition function, and realize microsimulation of individual trajectories in alignment with the Q-learning policy. Section \ref{sec:results_microsimulation} shows that the state transition probabilities estimated from the microsimulated individuals match well with those evaluated on real-data. 

\paragraph{Define rewards} We define deterministic reward in each month $t$, given state $s_t$ and action $a_t$, $\mathcal{R}(s_t, a_t) = - I(s_t, a_t) \times (1 + \alpha \times a_t)$,
where $I(s_t, a_t): \mathcal{S}\times\mathcal{A} \to \{0, 1\}$ is the severe COVID-19 infection status for the next month, and $\alpha$ represents the relative cost of receiving a booster. The next-month severe COVID-19 infection status $I(s_t, a_t)$ can be sampled using the probability generated by the fully trained RNN. The reward consists of two components: the predicted risk of severe infection in the next month and the potential side events of the booster. The parameter $\alpha$ quantifies the perceived harm of receiving a booster relative to the risk of a severe infection. As the Q-learning seeks a policy to maximize the expected cumulative reward, when $\alpha$ is small, it is likely to recommend boosters for all groups due to protective effect of vaccines against severe infections. Conversely, a large $\alpha$ may result in a policy that discourages booster administration, as the perceived harm outweighs the infection risk. In practice, the choice of $\alpha$ should reflect the relative importance placed on the risk of infection versus the harm of vaccination. We discuss our choice of $\alpha$ in Section \ref{sec:results_reward_eval} when presenting our results.

\section{Application on Learning COVID-19 Booster Policy}

In this section, we first present results of our microsimulations, showing that our RNN-based environment resembles real data well in terms of both marginal infection probabilities and conditional infection probabilities. Then, we present our Q-learning results, showing that the Q-table-based policy has an advantage over other policies in terms of rewards. We interpret the Q-table-based policy on whether and when to receive the booster on different groups of population when choosing selected vaccine costs. 

\subsection{Microsimulation to create environment}\label{sec:results_microsimulation}

\begin{figure}[t]
    \centering
    \small
    \spacingset{1}
    \begin{subfigure}{0.475\textwidth}
        \centering
        \includegraphics[scale = 0.5]{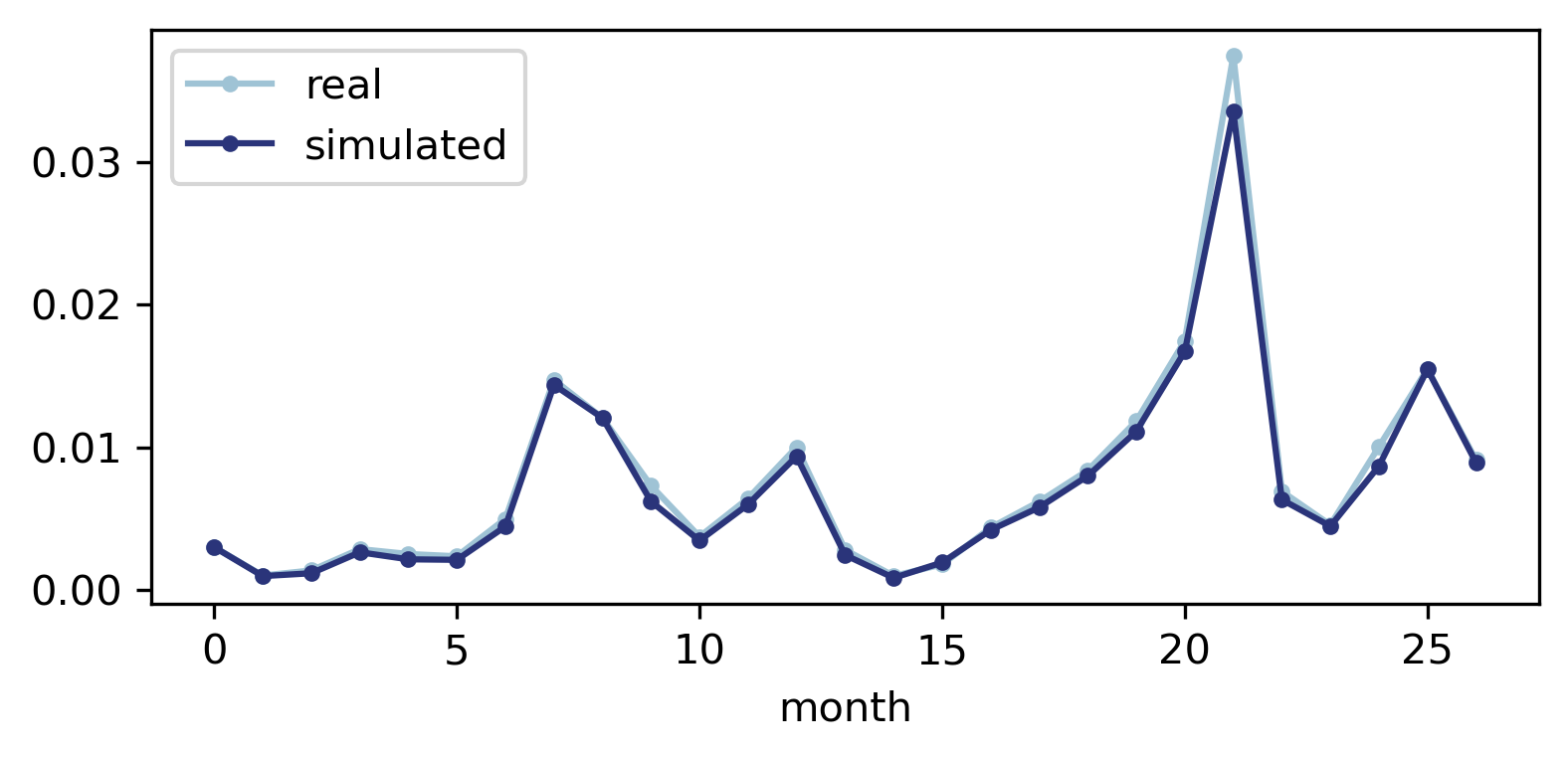}
        \caption{Marginal general infection rate by month.}
    \end{subfigure}
    \hfill
    \begin{subfigure}{0.475\textwidth}
        \centering
        \includegraphics[scale = 0.5]{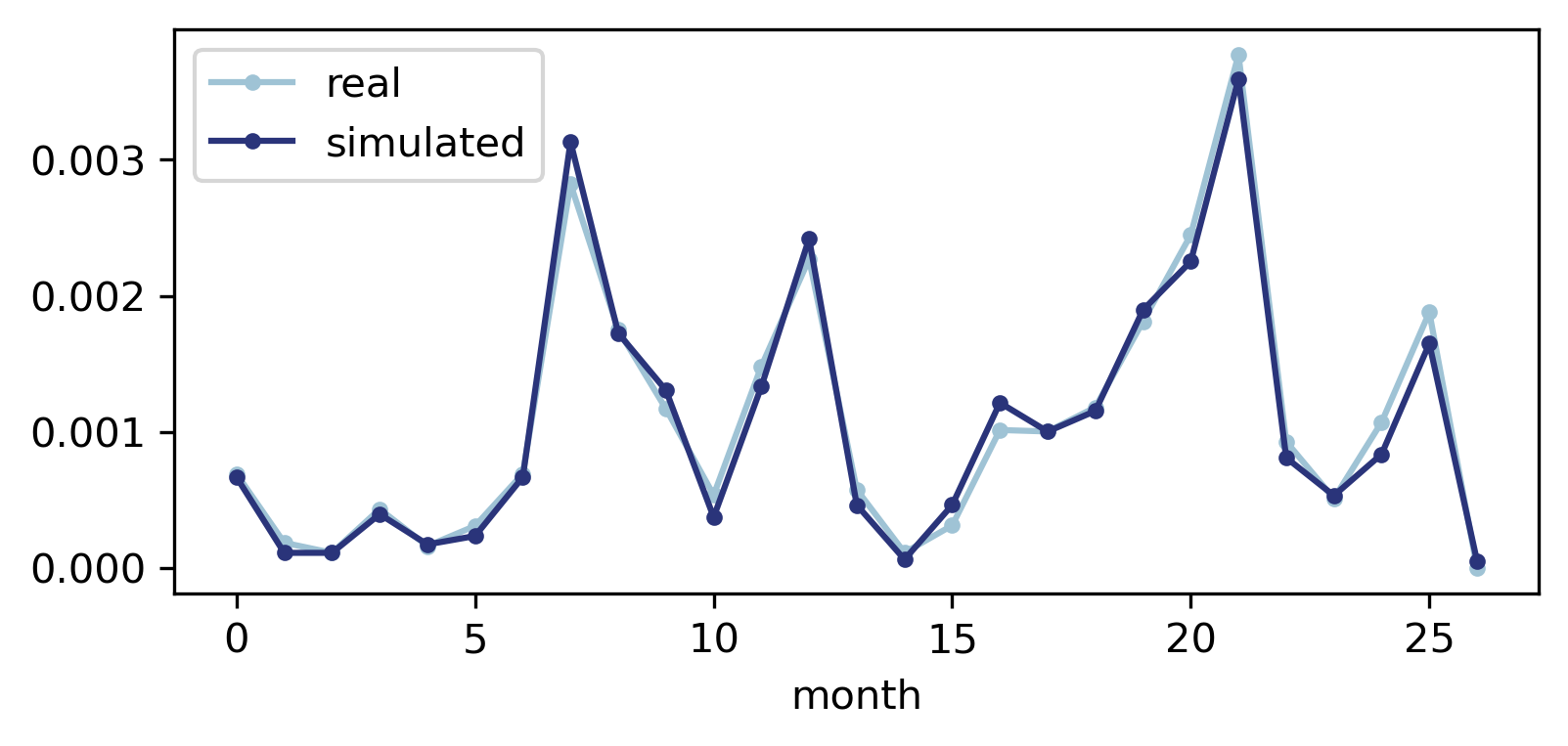}
        \caption{Marginal severe Infection rate by month.}
    \end{subfigure}
    \caption{Marginal general infection rate and marginal severe infection rate by month summarized from the simulated data and the real data.}
    \label{fig:inf_rate}
\end{figure}

We train the RNN using monthly data of the 81,000 patients starting from March 2020 to June 2022. We use an RNN with 2 stacked LSTM layer for training. Each LSTM layer has 128 hidden nodes with dropout rate 0.2 \citep{srivastava2014dropout}. During training, we use the Adam optimizer with learning rate $10^{-4}$ for 2,000 epochs \citep{kingma2014adam}. Sensitivity analysis shows that mild changes in hyperparameters do not change results significantly (see Supplementary Materials S2). To evaluate the RNN-based environment, we simulate a data sequence starting from the predictors at March 2020 for each of the 81,000 patients by the trained RNN. We summarize the severe infection rate and the general infection rate from the simulated data and compare them with the real EHR data. 

The simulated population marginal general infection rate and marginal severe infection rate over the 27 months are 7.25\textperthousand\;and 1.06\textperthousand\;respectively, which are close to the observed values in real data (7.91\textperthousand\;and 1.06\textperthousand\;respectively). Figure \ref{fig:inf_rate} shows the marginal general infection rate and marginal severe infection rate within the population at each month for the simulated data and the real data. The RNN-based environment fits very well both the marginal general infection rate and marginal severe infection rate within the population over the 27 months. The simulated marginal infection rates is very close to the real infection rates at each month. 

\begin{table}[t]
    \spacingset{1}
    \footnotesize
    \caption{Severe infection rate (\textperthousand) conditional on one variable: (a) age (b) number of previous COVID-19 vaccines (c) number of baseline hospital visits (d) comorbidity score.}
    \footnotesize
    \begin{subtable}{0.475\textwidth}
        \centering
        \caption{Age}
        \begin{tabular}{l c c}
        \hline\hline
        & \textbf{Simulated}   & \textbf{Real}\\
        \hline
        \textbf{Age [0, 18)} & 0.67 & 0.67 \\
        \textbf{Age [18, 30)} & 0.92 & 1.06 \\
        \textbf{Age [30, 50)} & 1.07 & 1.12 \\
        \textbf{Age [50, 65)} & 1.14 & 1.05 \\
        \textbf{Age 65+} & 1.26 & 1.26 \\
        \hline
        \bottomrule
        \end{tabular}
    \end{subtable}
    \hfill
    \begin{subtable}{0.475\textwidth}
        \centering
        \caption{Number of vaccines}
        \begin{tabular}{l c c}
        \hline\hline
        & \textbf{Simulated}   & \textbf{Real}\\
        \hline
        \textbf{numVax = 0} & 1.27 & 1.28 \\
        \textbf{numVax = 1} & 1.04 & 1.06 \\
        \textbf{numVax = 2} & 0.72 & 0.73 \\
        \textbf{numVax = 3} & 0.73 & 0.85 \\
        \textbf{numVax = 4} & 0.32 & 1.00 \\
        \hline
        \bottomrule
        \end{tabular}
    \end{subtable}
    \vfill
    \vspace{0.3cm}
    \begin{subtable}{0.475\textwidth}
        \centering
        \caption{Number of visits}
        \begin{tabular}{l c c}
        \hline\hline
        & \textbf{Simulated}   & \textbf{Real}\\
        \hline
        \textbf{numVisits [0, 5)} & 0.76 & 0.79 \\
        \textbf{numVisits [5, 10)} & 0.74 & 0.75 \\
        \textbf{numVisits [10, 20)} & 1.08 & 1.05 \\
        \textbf{numVisits [20, 50)} & 1.84 & 1.76 \\
        \textbf{numVisits 50+} & 2.84 & 3.28 \\
        \hline
        \bottomrule
        \end{tabular}
    \end{subtable}
    \hfill
    \begin{subtable}{0.475\textwidth}
        \centering
        \caption{Comorbidity}
        \begin{tabular}{l c c}
        \hline\hline
        & \textbf{Simulated}   & \textbf{Real}\\
        \hline
        \textbf{Comorbidity [0, 1)} & 0.76 & 0.77 \\
        \textbf{Comorbidity [1, 3)} & 1.43 & 1.39 \\
        \textbf{Comorbidity [3, 5)} & 2.18 & 2.24 \\
        \textbf{Comorbidity 5+} & 3.02 & 3.15 \\
        \hline
        \bottomrule
        \end{tabular}
    \end{subtable}
    \label{tab:conditional_infection_rate}
\end{table}

\begin{table}[t]
\spacingset{1}
\footnotesize
\caption{Severe infection rate (simulated/observed, \textperthousand) conditional on multiple variables: (a) Baseline immunosuppressant status and gender (b) COVID variant and race. }
\begin{subtable}{\textwidth}
    \centering
    \caption{Baseline immunosuppressant status and gender.}
    \begin{tabular}{l c c}
    \hline\hline
    & \textbf{imm\_baseline = 0} & \textbf{imm\_baseline = 1}\\
    \hline
    \textbf{gender = 0} & 0.95 / 0.95 & 1.53 / 1.51\\
    \textbf{gender = 1} & 1.01 / 1.05 & 2.02 / 1.85\\
    \hline
    \bottomrule
    \end{tabular}
\end{subtable}
\vfill
\vspace{0.3cm}
\begin{subtable}{\textwidth}
    \centering
    \caption{COVID variant and race}
    \begin{tabular}{l c c c}
    \hline\hline
    & \textbf{Variant None} & \textbf{Variant Delta} & \textbf{Variant Omicron}\\
    \hline
    \textbf{Race Caucasian} & 0.76 / 0.76 & 1.56 / 1.53 & 0.72 / 0.79\\
    \textbf{Race African American} & 1.71 / 1.61 & 4.60 / 4.57 & 1.30 / 1.55\\
    \textbf{Race Others} & 0.75 / 0.80 & 1.60 / 1.69 & 0.71 / 0.76\\
    \hline
    \bottomrule
    \end{tabular}
\end{subtable}
\label{tab:conditional_infection_rate_multiple_variables}
\end{table}

Table \ref{tab:conditional_infection_rate} shows the severe infection rate conditional on one variable and Table \ref{tab:conditional_infection_rate_multiple_variables} shows the severe infection rate conditional on multiple variables. The simulated conditional severe infection rates are very close to the rates observed in real data in most cases. Rarely, the simulated conditional severe infection rate has some difference with the observed values because there are limited data points within that category.

To assess whether the simulated decision process based on the trained RNN satisfies the Markov property, we use the test proposed by \citet{shi2020does} on multiple random subsets, and the low rejection rates (0\%–2\%) indicate that the Markov assumption is not violated (see Supplementary Materials S3 for details). These results suggest that the trained RNN provides a reasonable approximation of a Markovian environment, which justifies the use of (tabular) Q-learning to obtain an approximately optimal policy.

Overall, results show that the digital twin (RNN-based environment) is reliable for the online tabular Q-learning. The trained RNN can simulate data with very similar marginal and conditional infection rates with those in the real data. It well captures the relationship between the infection status and both the baseline and time-varying variables. 

\subsection{Booster policy learning}

We include three variables in state $\mathcal{S}$ in the online tabular Q-learning: age, baseline immunosuppressant status, and number of months to the second COVID-19 vaccination. Since we aim to learn a policy on whether to receive a COVID-19 booster at a specific month, we only consider subjects with at least two COVID-19 vaccinations. In this study, we focus on the policy of the first booster. 

\subsubsection{Reward evaluation}\label{sec:results_reward_eval}

We compare the population average rewards over months of the Q-table-based policy and three other policies: observed policy from data, all receiving a booster, and none receiving a booster. For the policy from data, we extract whether and when each subject received the first booster from the real data. For the policy of receiving a booster, we randomly pick a month between the 5th from the second vaccination and the last month of the study for each subject to receive a booster. For the policy of never receiving a booster, no one receives any booster at any month. We train the Q-table for 30 epochs and repeat the training 20 times with different random seeds. Supplementary Materials S1 shows that the tabular Q-learning algorithm converges before 30 epochs in our study. The discount factor $\gamma$ in the tabular Q-learning is fixed at 0.99. 

The vaccine cost $\alpha$ is an important hyperparameter that controls the belief of the relative harmness of a severe infection and the booster. One of the choice is to determine a reasonable range of vaccine cost based on mortality rates after a severe infection and the booster. We compute the mortality rate after 30 days among people who received the booster (0.04\%) and the mortality rate after 30 days among people who had a severe infection (1.05\%) from the real data. We use the ratio (0.04) of the two mortality rates for a proxy of the relative risk of a booster to a severe infection, so we primarily focus on the vaccine cost $\alpha$ around 0.04. 

\begin{figure}[t]
    \spacingset{1}
    \small
    \centering
    \includegraphics[scale = 0.3]{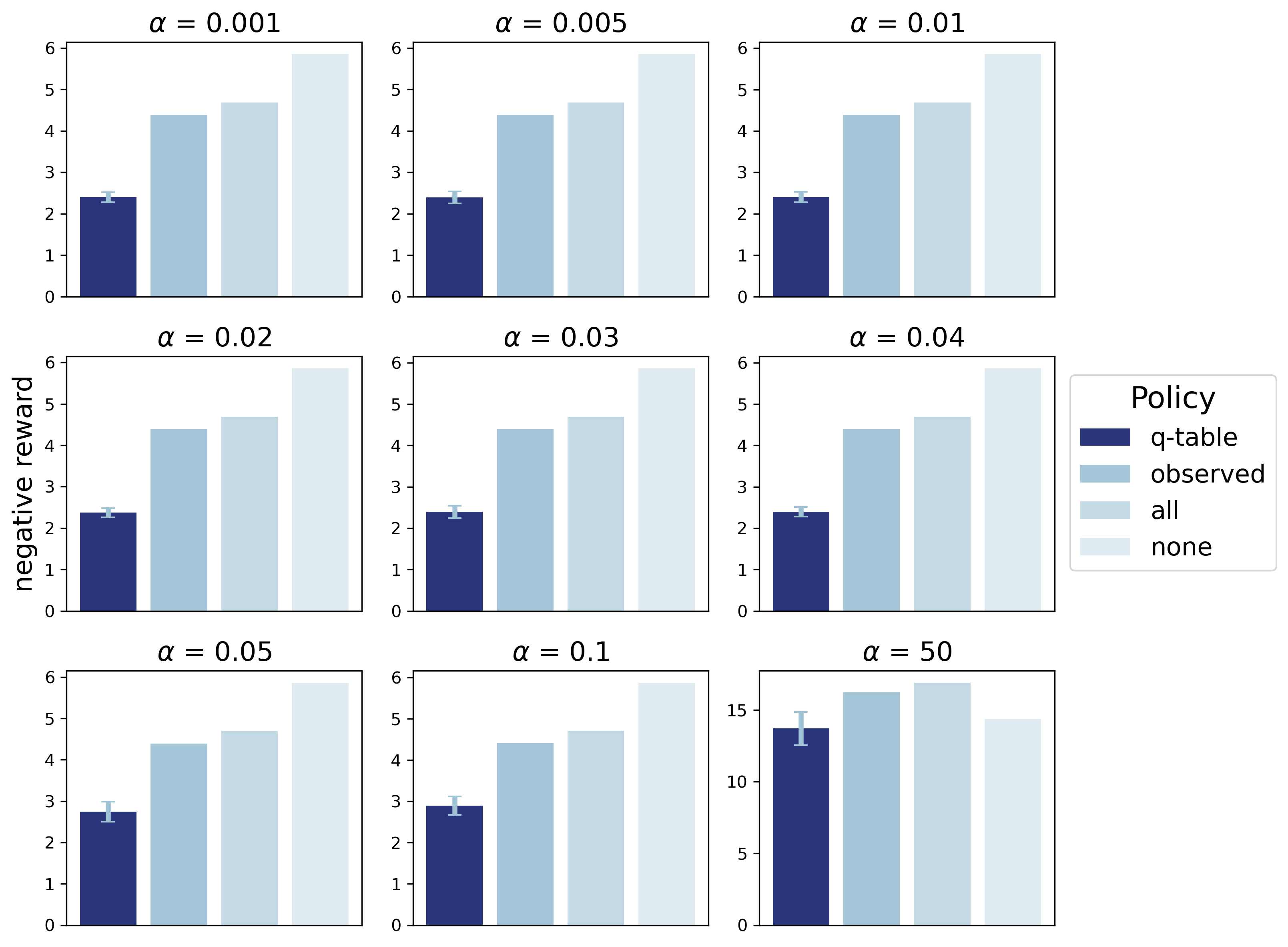}
    \caption{Negative rewards ($\times10^{-4}$) for the Q-table-based policy (q-table), the observed policy from data (observed), the policy of all receiving a booster (all), the policy of none receiving a booster (none). The error bar represents mean $\pm$ s.d. of the population average negative rewards over the 20 replicates for the Q-table-based policy. }
    \label{fig:reward}
\end{figure}

Figure \ref{fig:reward} shows the negative rewards (the lower the better) for the four policies and selected vaccine costs around 0.04. For all the vaccine costs, the Q-table-based policy consistently has the highest reward. The Q-table-based policy has a stable reward over the 20 replicates. When the vaccine cost increases (vaccine cost = 50), the Q-table-based policy gets close to the policy of never receiving a booster, as the large vaccine cost discourages it to receive a booster at any time point. It is worth-noting that the Q-table-based policy has a higher reward than the policy from the real data in all cases. This indicates the policy that was followed by the general population is sub-optimal and could have been improved by our Q-table-based policy.

Supplementary Materials S4 includes several sensitivity analyses, including varying different choices of $\gamma$, adding gender as an additional state, and refining age into more categories. Results show that our results are robust to the above variations.

In Section \ref{sec:comparison}, we provide the comparison between tabular Q-learning and deep Q-learning under different vaccine costs. Results show that deep Q-learning with various combinations of architectures and learning rates suffers from convergence difficulties and stability issues. This highlights the practical advantage of tabular Q-learning over deep Q-learning in this application. Another alternative is the Dyna-Q algorithm, which augments the standard Q-learning algorithm by additional planning steps \citep{sutton2018reinforcement}. We also apply the Dyna-Q algorithm, and its performance is comparable with standard Q-learning in our application. Supplementary Materials S5 includes additional results on the comparisons between Dyna-Q and standard tabular Q-learning. 

\subsubsection{Policy interpretation}

In Section \ref{sec:results_reward_eval}, we determine a reasonable vaccine cost is around 0.04 based on the mortality rates ratio after 30 days of a booster and a severe infection. Based on the 20 replicates, we obtain a confidence measure for receiving booster of the Q-table-based policy. We determine a group of people need to receive the booster if more than 10 replicates out of the 20 suggest so (i.e., confidence measure for receiving booster is bigger than $10/20=0.5$). 

When vaccine cost is set to 0.05, the Q-table policy recommends that adults aged 50-65 without baseline immunosuppressant use delay the booster until the 7th month after their second vaccination. In contrast, individuals of the same age group with baseline immunosuppressant use, as well as adults older than 65 without immunosuppressant use, are advised to receive the booster earlier, at 5-6 months. Among those aged 30-50, the policy suggests receiving the booster at 5-6 months if there is no immunosuppressant use, but postponing until the 7th month if there is baseline immunosuppressant use. When the vaccine cost rises to 0.1, the recommended policies remain unchanged, although the associated confidence measures for booster administration generally decrease. Conversely, when the vaccine cost decreases to 0.04, all adult age groups, regardless of immunosuppressant status, are advised to receive the booster 5-6 months after the second vaccination. Further decreasing the cost to 0.03 does not alter the recommended timing, but confidence measures increase to at least 0.9 for most groups. Across costs around 0.04, the confidence in recommending a booster at 5-6 months is generally higher for individuals with baseline immunosuppressant use compared to those without, and higher for adults over 65 compared to younger adults. Overall, these results suggest prioritization of booster vaccination for older adults and those with immunosuppressant use which aligns with prior findings of previous COVID-19 vaccine studies \citep{risk2022covid, risk2022comparative, shen2022efficacy}.

\section{Comparisons with Deep-Q Learning}\label{sec:comparison}

In this section, we compare tabular Q-learning and deep Q-learning for booster policy learning using the same microsimulation environment from Section \ref{sec:results_microsimulation}. Unlike tabular Q-learning, which stores the Q-function in a table, deep Q-learning approximates it with a neural network $Q_\theta(s,a)$ parameterized by $\theta$.

The deep Q-learning algorithm minimizes the temporal difference (TD) loss function, defined as $L(\theta) = \frac{1}{2}\left\{y - Q_\theta(s, a)\right\}^2$, where the target value $y=r(s,a)+\gamma\max_u Q_{\theta^{-}}(s', u)$ with $Q_{\theta^{-}}(s', u)$ representing the target network, which is periodically updated with the lagged parameters $\theta^{-}$ to stabilize training \citep{mnih2015human}. The parameters of the neural network are updated using gradient descent, following the rule $\theta\leftarrow\theta-\beta\nabla_\theta L(\theta)$ where $\nabla_\theta L(\theta)=-\delta\nabla_\theta Q_\theta(s, a)$ with approximated TD error $\delta=r(s,a)+\gamma\max_u Q_{\theta^{-}}(s', u)-Q_\theta(s,a)$. The gradient descent update rule can be written as $\theta\leftarrow\theta+\beta\delta\nabla_\theta Q_\theta(s, a)$, which mirrors the tabular Q-learning update rule in (\ref{eq:tabular_q_update}), i.e., $Q(s, a)\leftarrow Q(s,a) + \beta\delta^*$, where $\delta^*=r(s,a)+\gamma\max_u Q(s', u)-Q(s,a)$ represents the TD error. The TD error $\delta$ in deep Q-learning is an approximation of the true TD error, as it relies on the maximum Q-values calculated using the lagged parameters of the target network. Consequently, while tabular Q-learning offers convergence guarantees under standard conditions, deep Q-learning does not share the same theoretical guarantees of optimality due to the inherent approximation and instability introduced by function approximation.

\begin{figure}[t!]
    \spacingset{1}
    \small
    \centering
    \includegraphics[scale = 0.25]{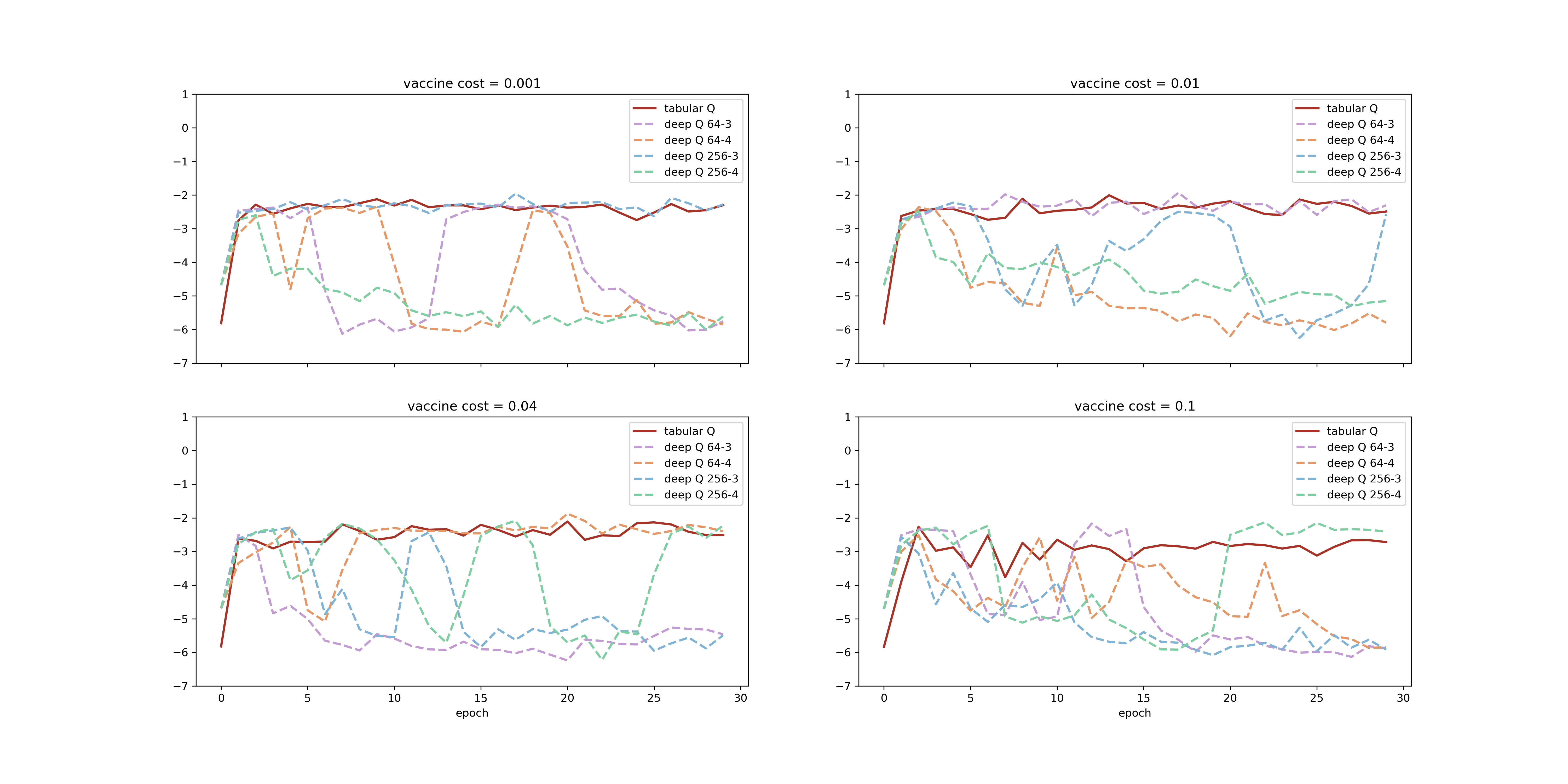}
    \caption{Comparison between tabular Q-learning and deep Q-learning with different architectures: the average reward evaluated over individuals along 30 epochs for different vaccine costs. }
    \label{fig:epoch_reward_deep}
\end{figure}

In Figure \ref{fig:epoch_reward_deep}, we present the average reward over individuals along 30 epochs for deep Q-learning, considering various neural network architectures and learning rates. For comparison, we also show the average reward for tabular Q-learning over the same number of training epochs. The deep Q-learning models use networks with two hidden layers, each containing either 64 or 256 nodes. For each architecture, we consider two learning rates, $10^{-3}$ and $10^{-4}$, with the Adam optimizer \citep{kingma2014adam}. For instance, in Figure \ref{fig:epoch_reward_deep}, the line with label ``deep Q 64-4'' corresponds to a Q-network of two hidden layers with 64 nodes each, trained by the Adam optimizer with learning rates $10^{-4}$.

Figure \ref{fig:epoch_reward_deep} suggests that deep Q-learning suffers from convergence difficulties and stability issues for all combinations of architectures and learning rates. The Q-network with two layers of 256 nodes and trained with learning rate $10^{-3}$ (deep Q 256-3) converges when vaccine cost is $0.001$, but does not converge for other vaccine costs. Similarly, the Q-network with two layers of 64 nodes, trained with learning rate $10^{-3}$ (deep Q 64-3), converges only when vaccine cost is $0.01$. The Q-network with two layers of 64 nodes and trained with learning rate $10^{-4}$ (deep Q 64-4) converges only when vaccine cost is $0.04$. The same Q-network architecture trained with the same learning rate exhibits highly variable performance depending on the vaccine cost. Even when deep Q-learning converges, its has similar reward compared to that of tabular Q-learning. In contrast, tabular Q-learning demonstrates consistent convergence and robust performance across all scenarios. This highlights the practical advantage of tabular Q-learning over deep Q-learning in public health applications, where the state and action spaces are often discrete and relatively small. Given the challenges associated with deep Q-learning, especially in terms of stability and convergence, tabular Q-learning remains a reliable and efficient choice when the problem structure allows for it.

\section{Discussion and Conclusion}

In this paper, we propose a novel framework combining tabular Q-learning with an RNN-based environment simulator to optimize COVID-19 booster vaccination policies. The proposed approach addresses key challenges in vaccine policy development, including the limitations of clinical trials and ethical concerns on need of direct interactions of the Reinforcement Learning (RL) algorithms with the real world. By utilizing an RNN, we successfully create a digital twin of the infection dynamics for the target population, which models the temporal relationships of COVID-19 infections and vaccination status, generating simulated data that reflects real-world dynamics. The policy learned through our method outperforms the currently observed practices of COVID-19 booster vaccination, indicating its potential to enhance vaccine deployment and reduce infection rates.

A valid application of our framework relies on the Markov assumption. The RNN used in the microsimulation is capable of modeling both Markovian and non-Markovian transitions. We test whether the transition dynamics induced by the RNN satisfy the Markov property \citep{shi2020does}. Applied to the simulated data, this test indicates no evidence against the Markov assumption, which justifies the use of tabular Q-learning in our application. However, this justification is not at odds with the use of an RNN. While the RNN is capable of encoding non-Markovian dependence, it does not impose such a structure if the data do not support it. In applications where the Markov assumption is violated, one can mitigate non-Markovianity by augmenting the state with relevant past observation–action pairs (with the history length potentially determined adaptively, e.g., by \citet{shi2020does}) or by incorporating learned RNN representations, such as the final hidden layer, into the state definition. Such augmentations, however, would result in a state space that mixes discrete and continuous components, in which case a straightforward policy table is no longer directly available. Nonetheless, interpretable policy summaries can still be obtained for clinically relevant groups by marginalizing over the continuous state variables.

Our framework offers several advantages. First, the RNN-generated simulated data enables continuous exploration of potential policies without ethical concerns. This allows us to conduct extensive policy evaluations without requiring real-world interventions, avoiding the risks of harmful or suboptimal decisions. Second, by employing tabular Q-learning, we provide an interpretable and clear policy table, allowing policymakers to easily understand and implement optimal vaccination strategies. While Deep Q-learning has been widely applied in healthcare for its flexibility in large and continuous state spaces, it suffers from convergence difficulties and stability issues in our applications. This instability highlights the value of tabular Q-learning, which, while simpler, offers more reliable and interpretable outcomes for public health problems where states and actions are discrete.

This research demonstrates the effectiveness of RL in public health policy development and presents a scalable solution for future pandemics or vaccine rollouts. Although we focus on COVID-19 booster policymaking as a case study, our approach is broadly applicable to problems requiring decision-making for different populations. Beyond vaccines for future pandemics and other public health interventions (e.g., smoking cessation strategies), a further example is cancer screening, where recommendations vary according to factors such as individual risk status, age, and family history \citep{ACS2023}.

\if1\blind
{
  \vspace{0.7cm}
  \paragraph{Acknowledgement} The authors thank the University of Michigan Data Office for assistance with data extraction from electronic medical records. Dr. Zhao's research is supported by the National Institute of Allergy and Infectious Diseases of the National Institutes of Health under award number R01AI158543. Dr. Kang’s research was partially supported by National Institute of Health grants R01DA048993 and R01MH105561 and the National Science Foundation grant IIS-2123777. The content is solely the responsibility of the authors and does not necessarily represent the official views of the National Institutes of Health.
} \fi

\clearpage
\spacingset{1.2}
\bibliographystyle{apalike}
\bibliography{ref}

\end{document}


\spacingset{1.3}
\maketitle

\section*{S1. Rewards of tabular Q-learning over epochs}

We provide the plot of mean running reward over 30 epochs of different vaccine costs for tabular Q-learning in Figure S1. The reward is averaged over all individuals and all time points. The initial policy is never receiving at any time point for all vaccine costs. For all vaccine costs, the mean reward converges before 30 epochs, showing that training the Q-learning algorithm for 30 epochs is sufficient. 

\begin{figure}[b!]
    \centering
    \includegraphics[scale = 0.4]{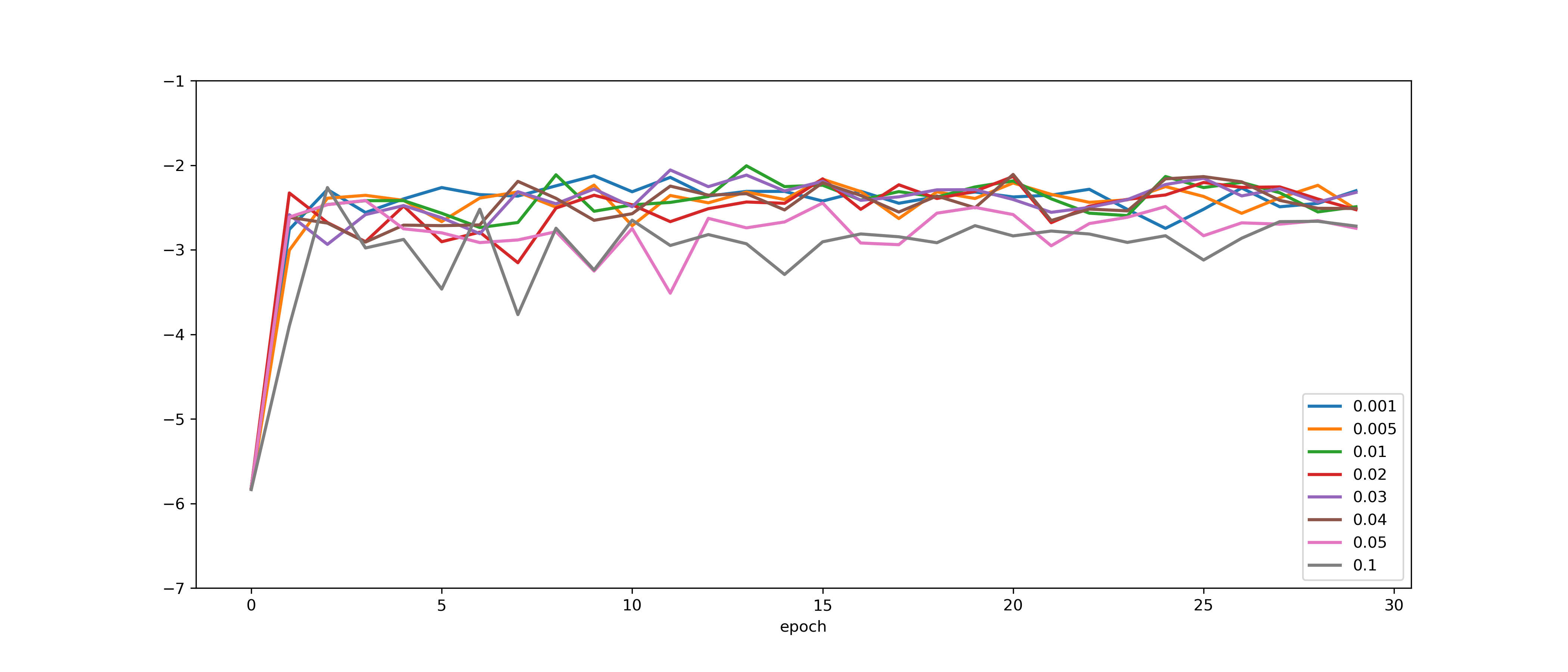}
    \caption*{Figure S1. The mean running reward ($\times 10^{-4}$) over 30 epochs of different vaccine costs for tabular Q-learning. }
\end{figure}

\section*{S2. Sensitivity analysis for RNN}

We include a sensitivity analysis for RNN training in Table S1. We check the mean absolute error (MAE, in 1e-2) for general infection rate and severe infection rate over time across different RNN training parameter settings. The first row is corresponding to the RNN architecture used in the main text. We vary the number of stacked LSTM layers, number of neurons in each layer, number of epochs and learning rate. Each setting is based on scaling one parameter by 0.5 or 2 while keeping others the same as current value. The MAE stays stable across settings.

\begin{table}[t]
    \small
    \centering
    \begin{tabular}{c c c c c c}
    \hline\hline
    \textbf{\# Layers} & \textbf{\# Neurons} & \textbf{Epochs} & \textbf{Learning Rate} & \textbf{MAE (General)} & \textbf{MAE (Severe)} \\
    \hline
    2 & 128 & 2000 & $1\times10^{-4}$   & 0.04 & 0.011 \\
    2 & 128 & 2000 & $2\times10^{-4}$   & 0.04 & 0.015 \\
    2 & 128 & 2000 & $5\times10^{-5}$   & 0.04 & 0.011 \\
    2 & 128 & 4000 & $1\times10^{-4}$   & 0.04 & 0.015 \\
    1 & 128 & 2000 & $1\times10^{-4}$   & 0.05 & 0.013 \\
    2 & 128 & 1000 & $1\times10^{-4}$   & 0.05 & 0.014 \\
    2 & 256 & 2000 & $1\times10^{-4}$   & 0.05 & 0.012 \\
    2 & 64  & 2000 & $1\times10^{-4}$   & 0.05 & 0.013 \\
    4 & 128 & 2000 & $1\times10^{-4}$   & 0.05 & 0.012 \\
    \hline
    \bottomrule
    \end{tabular}
    \caption*{Table S1. The mean absolute error (MAE, in $10^{-2}$) for general infection rate and severe infection rate over time across different RNN training parameter settings. The first row is corresponding to the RNN architecture used in the main text. We vary number of stacked LSTM layers, number of neurons in each layer, number of epochs and learning rate. Each setting is based on scaling one parameter by 0.5 or 2 while keeping others the same as current value.}
\end{table}

\section*{S3. Test for Markov Property}

To assess whether the simulated decision process based on the trained RNN satisfies the Markov property, we employ the test proposed by \citet{shi2020does}, which evaluates whether future outcomes depend on the current state-action pair only, or also on additional past history. Our state definition for this test includes all relevant time-varying variables used in Q-learning, such as vaccination history and severe infection status, as well as the action (vaccinate or not). Time-invariant covariates such as baseline age and immunosuppressant usage are excluded from the test since they remain constant across time and their influence on the outcome can be viewed as deterministic conditioning. To implement the test, we simulate data using our RNN-based environment model and conduct the Markov test on randomly sampled subsets due to computational constraints. Specifically, we sample subsets of size 300, 500, and 1000 from a dataset of 81,000 subjects. For each sample size, we run the test on multiple random subsets (100 for sizes 300 and 500, and 50 for size 1000). We setn\_{trees} = 200, B = 200 and other arguments as package default for all tests. All tests We compute the proportion of subsets in which the null hypothesis (the Markov property holds) is rejected at the 5\% level. The rejection rates were 0\%, 0\%, and 2\% for sample sizes of 300, 500, and 1000 respectively. These results suggest that the trained RNN provides a reasonable approximation of a Markovian environment, which justifies the use of (tabular) Q-learning to obtain an approximately optimal policy.

\section*{S4. Sensitivity Analysis for Tabular Q-learning}

\paragraph{Varying discount factor $\gamma$} We conduct sensitivity analysis on a range of different discount factors from 0.5 to 0.999. Figure S2 shows the bar plot for the negative average reward (the lower the better) of policies based on different $\gamma$’s. We repeat each setting on 20 replicates. Error bars represent the standard deviation estimated based on the 20 replicates. The results show that the Q-learning performance does not vary significantly when $\gamma$ is over 0.7. Therefore, the model performance is not sensitive to the choice of $\gamma$ in this study. 

\begin{figure}[t]
    \centering
    \includegraphics[width=0.4\linewidth]{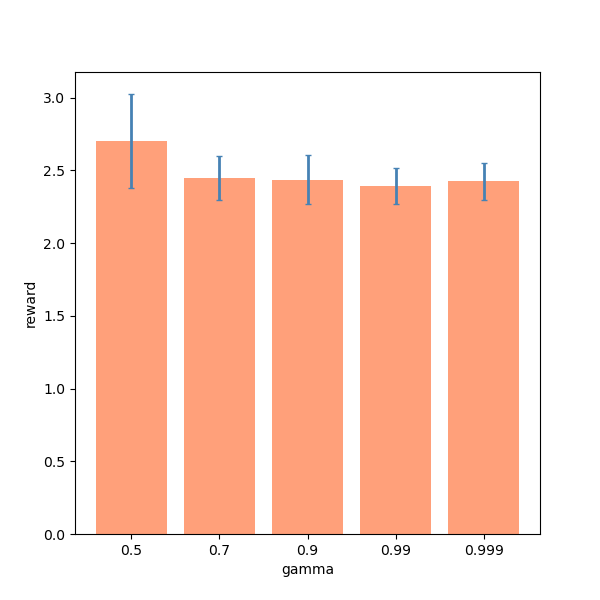}
    \caption*{Figure S2. Bar plot for the negative average reward (the lower the better) of policies based on different discount factor $\gamma$'s. Each setting is repeated on 20 replicates. Height of the bar represents the 20 replicate average and error bars represent the standard deviation estimated based on the 20 replicates. }
\end{figure}

\paragraph{Adding gender as an additional state in policy learning} We included gender as a state in the policy learning. Figure S3 shows the bar plot for the negative average reward (the lower the better) of policies on three settings. “Paper” refers to the setting reported in the main text and “add gender” refers to the setting where we add gender as an additional state. The results show that the Q-learning performance does not change significantly across settings. 

\begin{figure}[t]
    \centering
    \includegraphics[width=0.4\linewidth]{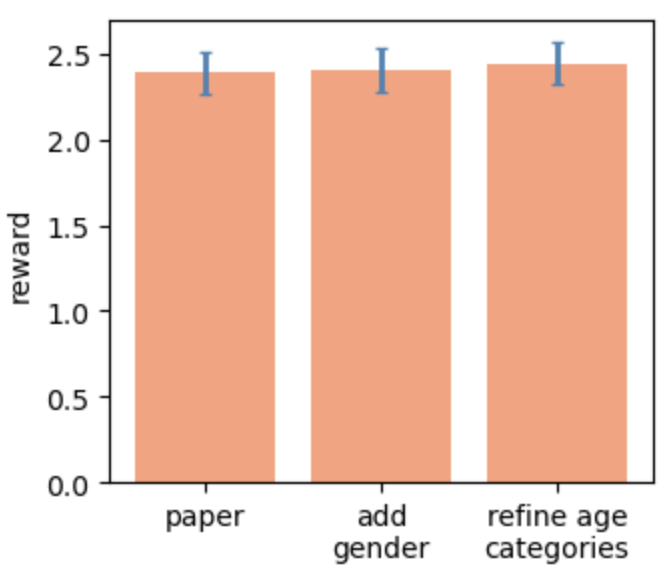}
    \caption*{Figure S3. Bar plot for the negative average reward (the lower the better) of policies for three settings. “Paper” refers to the setting reported in the main text, “add gender” refers to the setting where we add gender as an additional state, and “refine age categories” refers to the setting where we refine age into more categories. Each setting is repeated on 20 replicates. Height of the bar represents the 20 replicate average and error bars represent the standard deviation estimated based on the 20 replicates. }
\end{figure}

\begin{figure}[t]
    \centering
    \includegraphics[width=0.9\linewidth]{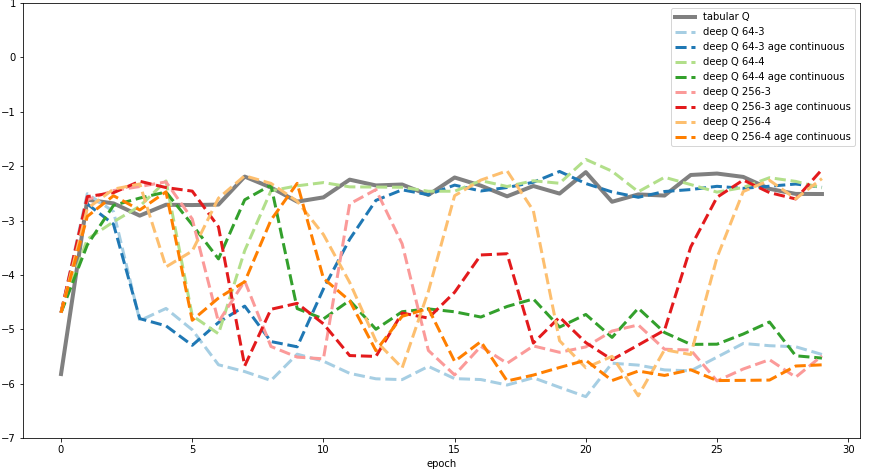}
    \caption*{Figure S4. Comparison between tabular Q-learning and deep Q-learning with different architectures: the average reward evaluated over individuals along 30 epochs for different vaccine costs. Deep Q-learning labeled without ``age continuous'' corresponds to those reported in Section 4. Deep Q-learning labeled with ``age continuous'' corresponds to the same architecture of Q-network reported in Section 4 but treat age as a continuous variable. }
\end{figure}

\paragraph{Refining age categories in policy learning} We refine the age categories to 0-12, 13-18, 19-25, 26-30, 31-50, 51-65, 66-80 and 81-100. Figure S3 shows the bar plot for the negative average reward (the lower the better) of policies based on the refined age category (refine age category) and the original age category (paper). The results show that the Q-learning performance does not change significantly across settings. 

\paragraph{Treating age as a continuous variable} Figure S4 shows the trace plot of reward for deep Q-learning where age is treated as a continious variable, compared to deep Q-learning where age is categorized. The results show that categrization does not undermine the performance in policy evaluation.

\section*{S5. Comparison between Dyna-Q and standard tabular Q-learning}

We apply the Dyna-Q algorithm which includes additional planning steps to the standard Q learning \citep{sutton2018reinforcement}. The hyperparameters for the Dyna-Q algorithm are set to be the same as those in standard tabular Q-learning algorithm for a fair comparison. For additional hyperparameters in the Dyna-Q algorithm, we set 10 planning steps after each interaction with the environment, and maintain a model buffer including 200,000 transitions where the Dyna-Q algorithm sample from in the planning steps.

Figure S5 shows the comparison between the Dyna-Q algorithm and the standard tabular Q-learning algorithm. We run both algorithms in the booster application on 20 replicates. Height of the bar represents the average negative reward over the 20 replicates and error bar represents the estimated standard deviation. In this application, we see the two methods have comparable performances.

\begin{figure}
    \centering
    \includegraphics[width=0.5\linewidth]{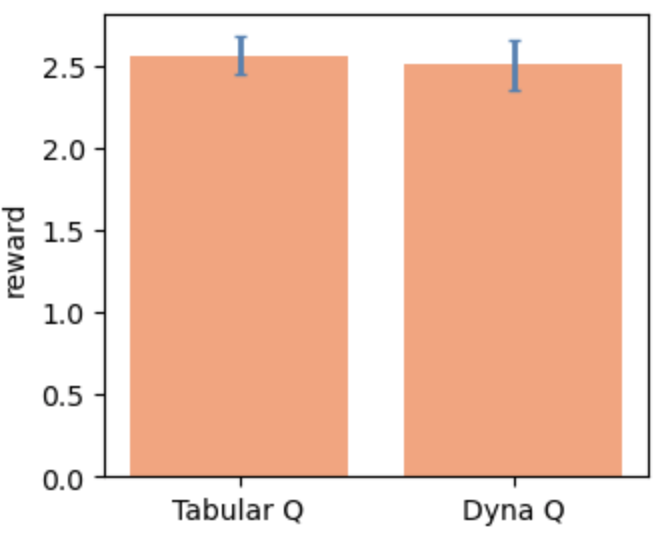}
    \caption*{Figure S5. Bar plot for the negative average reward (the lower the better) of policies based on the Dyna-Q algorithm and the standard tabular Q-learning algorithm. Each is repeated on 20 replicates. Height of the bar represents the 20 replicate average and error bars represent the standard deviation estimated based on the 20 replicates. }
\end{figure}

\section*{S6. Algorithm for online tabular Q-learning with the RNN-based environment in booster policy development}

Algorithm S1 outlines the steps of online tabular Q-learning with the RNN-based environment in our application of booster policy development. 

\begin{algorithm}[t]
\spacingset{1}
\begin{algorithmic}[1]
    \STATE Initialize Q-table $q(s, a) = 0$ for all $s$ and $a$
    \STATE Initialize $\epsilon=0.5$ and $\beta=0.001$, fix $\gamma=0.99$, $\lambda_\epsilon=0.99$, $\lambda_\beta=0.998$, and $k_\epsilon=k_\beta=5000$, which are commonly used values \citep{sutton2018reinforcement}
    \FOR{individual $i$ in $1:n$}
        \STATE Extract age, immunosuppressant usage, and the month of the second vaccination $T_i$ for individual $i$ from EHR data
        \STATE Set months since the last vaccine to 0 for individual $i$
        \STATE Initialize state $s_0$ with age, immunosuppressant usage and months since the last vaccine for individual $i$
        \FOR{time $t$ in $T_i+1:T$}
          \IF{$t - T_i\leq4$}
            \STATE Select $a_t=0$ by following the CDC guideline
          \ELSE
            \STATE Select random action $a_t$ with probability $\epsilon$, otherwise select $a_t=\arg\max_a q(s_t, a)$
          \ENDIF
          \STATE Obtain $r_t$, $s_{t+1}$ and $I(s_t, a_t)$ from the RNN micro-simulated environment; collect transition $(s_t, a_t, r_t, s_{t+1})$
          \STATE Update $q(s_t, a_t)\leftarrow q(s_t, a_t) + \beta\Big(r_t + \gamma\max_a q(s_{t+1}, a) - q(s_t, a_t)\Big)$
          \IF{$I_{t+1}$ = 1}
            \STATE Terminate simulation for individual $i$, continue to the next individual
          \ENDIF
          \STATE Decay $\epsilon\leftarrow\lambda_\epsilon\epsilon$ every $k_\epsilon$ steps, decay $\beta\leftarrow\lambda_\beta\beta$ every $b_\beta$ steps
        \ENDFOR
    \ENDFOR
\end{algorithmic}
\caption*{Algorithm S1. Online tabular Q-learning in RNN based environment simulator}
\end{algorithm}

\clearpage
\bibliographystyle{apalike}
\bibliography{ref}